\documentclass[12pt]{article}
\usepackage{graphicx}
\usepackage{subcaption}

\newcommand\pubnumber{SNSN-323-63}
\newcommand\pubdate{\today}

\def\qm{$^a$Queen Mary University of London}
\def\rmtre{$^b$INFN, Sezione di Roma Tre}
\def\ox{$^c$University of Oxford}
\def\rmuno{$^d$INFN, Sezione di Roma}
\def\rmtrebis{$^e$INFN, Sezione di Roma Tre, and Universit{\`a} di Roma Tre}
\def\sissa{$^f$SISSA-ISAS}
\def\ge{$^g$Universit\`a di Genova and INFN}
\def\cal{$^h$California Institute of Technology}
\def\lyon{$^i$IPNL-IN2P3 Lyon}
\def\paris{$^j$IN2P3-CNRS et Universit\'e de Paris-Sud}
\def\bo{$^k$INFN, Sezione di Bologna}

\def\speaker{\footnote{Speaker}}

\newcommand{\dmd}{\Delta m_d}
\newcommand{\dms}{\Delta m_s}

\newcommand{\rhob}{\bar \rho}
\newcommand{\etab}{\bar \eta}
\newcommand{\vubsvcb}{\left | V_{ub}/V_{cb}  \right |}

\def\utfit{{\bf{U}}\kern-.20em{\bf{T}}\kern-.15em{\it{fit}}\@}
\def\butfit{{\bf{U}}\kern-.18em{\bf{T}}\kern-.10em{\it{fit}}\@}

\def\Title#1{\begin{center} {\Large #1 } \end{center}}
\def\Author#1{\begin{center}{ \sc #1} \end{center}}
\def\Address#1{\begin{center}{\vspace*{-0.2cm} \it #1} \end{center} \vspace*{-0.2cm}}

\newcommand\pubblock{\rightline{\begin{tabular}{l} \pubnumber\\
         \pubdate  \end{tabular}}}
\newenvironment{Abstract}{\begin{quotation}  }{\end{quotation}}
\newenvironment{Presented}{\begin{quotation} \begin{center} 
             PRESENTED AT\end{center}\bigskip 
      \begin{center}\begin{large}}{\end{large}\end{center} \end{quotation}}

\def\beq{\begin{equation}}
\def\eeq#1{\label{#1}\end{equation}}
\def\eeqn{\end{equation}}

\def\beqa{\begin{eqnarray}}
\def\eeqa#1{\label{#1}\end{eqnarray}}
\def\eeqan{\end{eqnarray}}

\let\bar=\overbar

\def\Dslash{\not{\hbox{\kern-4pt $D$}}}
\def\dslash{\not{\hbox{\kern-2pt $\del$}}}

\def\BR{\mbox{\rm BR}}

\def\msb{{\bar{\ssstyle M \kern -1pt S}}}

\begin{document}
\begin{titlepage}
\pubblock

\vfill
\Title{Standard Model updates and new physics analysis with the Unitarity
Triangle fit}
\vfill

\Author{A.~Bevan$^a$, M.~Bona$^{a,}$\speaker, M.~Ciuchini$^b$, D.~Derkach$^c$, E.~Franco$^d$,
V.~Lubicz$^e$, G.~Martinelli$^{d,f}$, F.~Parodi$^g$, M.~Pierini$^h$, C.~Schiavi$^g$,
L.~Silvestrini$^d$, V.~Sordini$^i$, A.~Stocchi$^j$, C.~Tarantino$^e$, and V.~Vagnoni$^k$\\
[3mm]
\utfit\ Collaboration}

\vspace*{0.5cm}
\Address{\qm}       
\Address{\rmtre}    
\Address{\ox}       
\Address{\rmuno}    
\Address{\rmtrebis} 
\Address{\sissa}    
\Address{\ge}       
\Address{\cal}      
\Address{\lyon}     
\Address{\paris}    
\Address{\bo}       
\vspace*{0.3cm}

\vfill
\begin{Abstract}
We present here the update of the Unitarity Triangle (UT) analysis
performed by the \utfit~Collaboration within the Standard Model (SM)
and beyond. Continuously updated flavour results contribute to
improving the precision of several constraints and through the global fit
of the CKM parameters and the SM predictions.
We also extend the UT analysis to investigate new physics (NP)
effects on $\Delta F=2$ processes.
Finally, based on the NP constraints, we derive upper bounds
on the coefficients of the most general $\Delta F=2$ effective
Hamiltonian. These upper bounds can be translated into lower
bounds on the scale of NP that contributes to these
low-energy effective interactions.
\end{Abstract}
\vfill
\begin{Presented}
8th International Workshop on the CKM Unitarity Triangle (CKM 2014),
Vienna, Austria, September 8-12, 2014
\end{Presented}
\vfill
\end{titlepage}
\def\thefootnote{\fnsymbol{footnote}}
\setcounter{footnote}{0}

\section{Introduction}
One of the main tasks of Flavor Physics is an accurate determination
of the parameters of the Cabibbo-Kobayashi-Maskawa (CKM) matrix.
It represents a crucial test of the SM and, moreover, improving the
accuracy on the CKM parameters is at the heart of many searches for
NP, where small NP effects are looked for.

\section{Unitarity Triangle Analysis in the SM}
\label{sec:sm}

The Unitarity Triangle (UT) analysis presented here is performed by the
\utfit~Collaboration following the method described in
Refs~\cite{Ciuchini:2000de,Bona:2005vz}. From a Bayesian global fit
the CKM parameters $\rhob$ and $\etab$ from the Wolfenstein
parameterisation are obtained exploiting a plethora of flavour
measurements, both theoretical and experimental.
The basic constraints are: $\vubsvcb$ from semileptonic
$B$ decays, $\dmd$ and $\dms$ from $B^0_{d,s}$ oscillations,
$\varepsilon_K$ from $K$ mixing, $\alpha$ from charmless hadronic
$B$ decays, $\gamma$ from charm hadronic $B$ decays,
and $\sin2\beta$ from $B^0\to J/\psi K^0$ decays.
The complete set of numerical values used as inputs can be found at the URL
{\tt http://www.utfit.org}~\footnote{The results presented here are
an update on the ``Summer 2014'' analysis and they will appear in the
web-page as ``Winter 2015''.}.
The experimental measurements are mostly taken
from Ref.~\cite{hfag12}, while the non-perturbative QCD parameters come
from the most recent lattice QCD averages~\cite{flag13}.

\begin{figure}[!tb]
\centering
\vspace*{-1.6cm}
\hspace*{-0.6cm}
\includegraphics[width=0.36\linewidth]{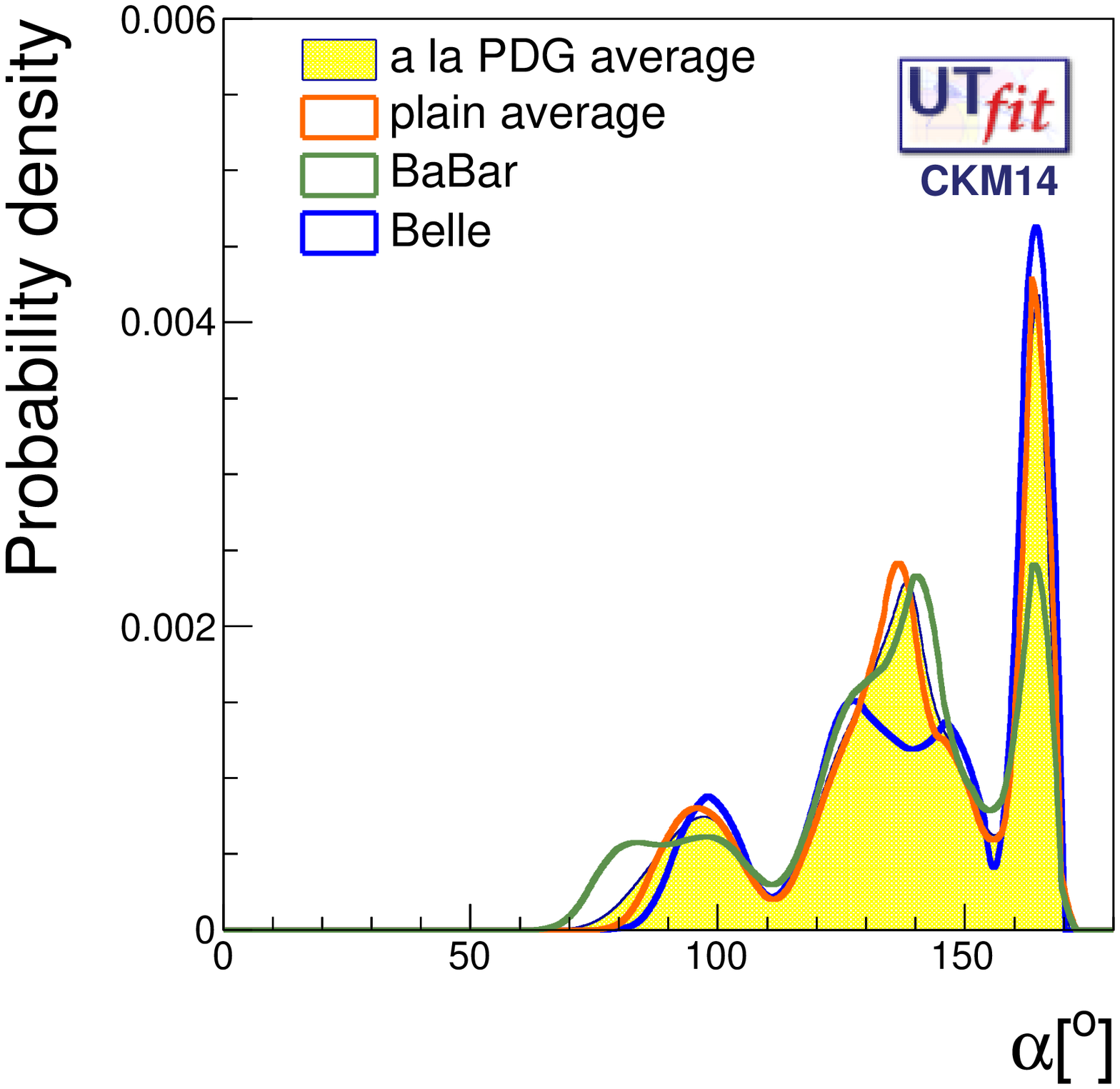}
\hspace*{-0.6cm}
\includegraphics[width=0.36\linewidth]{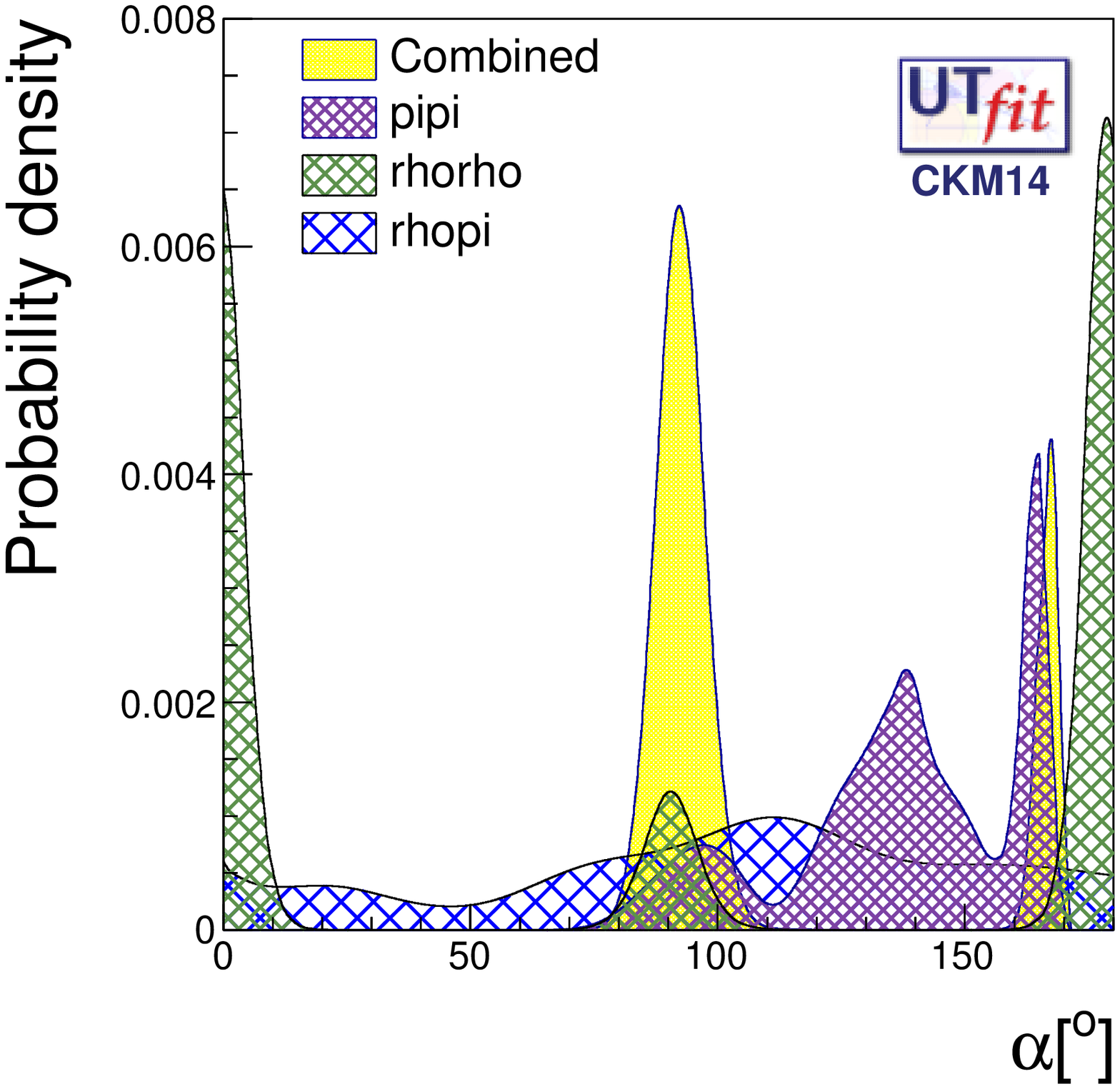}
\hspace*{-0.6cm}
\includegraphics[width=0.36\linewidth]{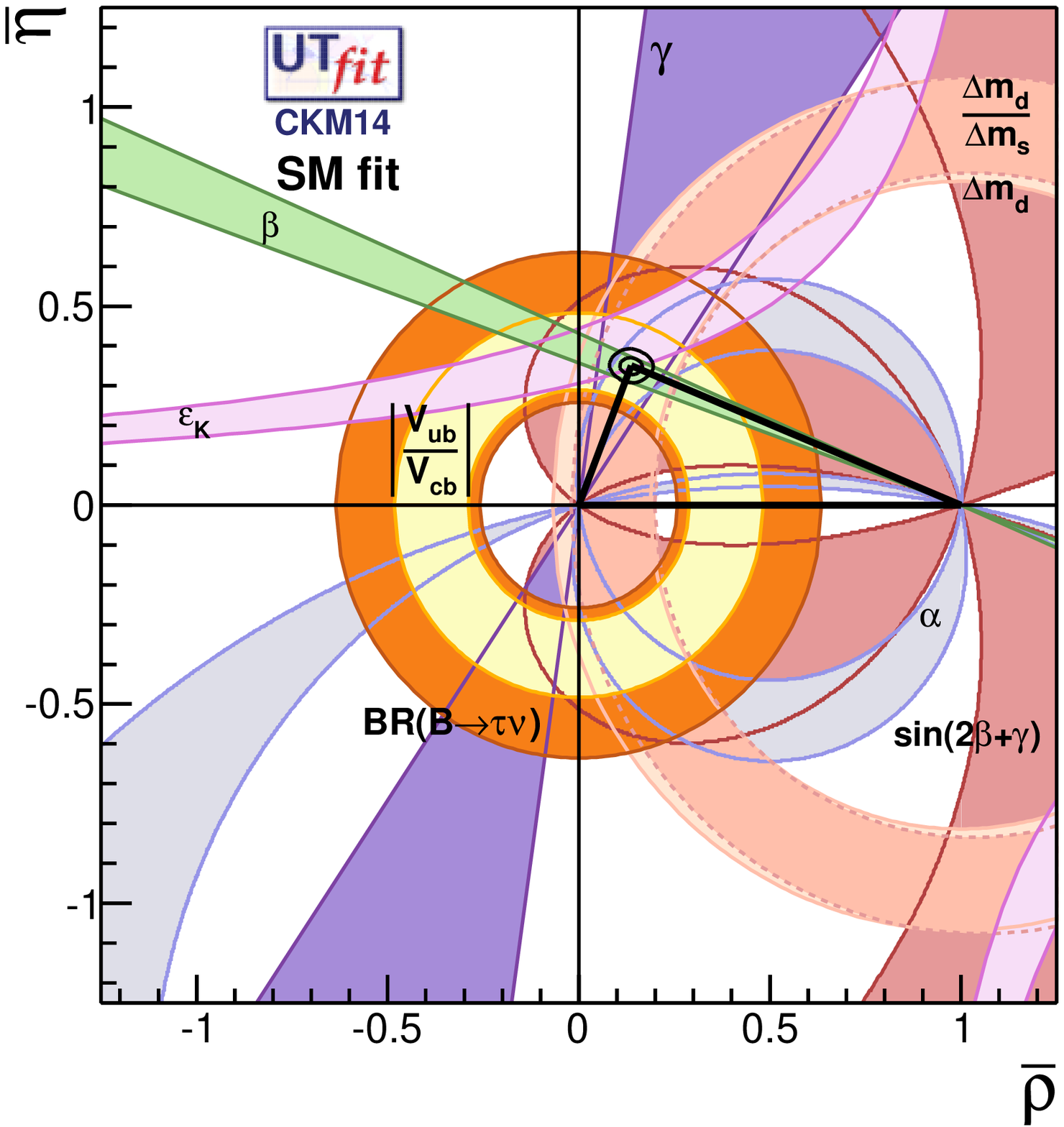}
\hspace*{-0.6cm}
\caption{{\it{Left}}: angle $\alpha$ (or $\phi_2$) extracted from the
$\pi\pi$ isospin analysis with different values of the $B\to\pi^0\pi^0$
BR. {\it{Middle}}: angle $\alpha$ extracted from the combination of
the different charmless $B$ decay systems. Both plots show the
one-dimensional p.d.f. for the given variable.
{\it{Right}}: $\rhob-\etab$ plane showing the result of the SM fit.
The black contours display the 68\% and 95\% probability
regions selected by the given global fit. The 95\% probability
regions selected by the single constraints are also shown.}
\label{fig:sm}
\end{figure}

Consider the angle $\alpha$ of the CKM triangle: it is extracted from
charmless hadronic $B$ decays with the method described in~\cite{UTalpha}.
For this report, we updated the $\alpha$ extraction with the latest input
values, in particular using the $B^0\to\pi^0\pi^0$ BR with
the new average $(1.15 \pm 0.41)\:10^{-6}$, including the latest Belle
result~\cite{bellepi0pi0}.
The left plot in Figure~\ref{fig:sm} shows the effect on the 
$B^0\to\pi^0\pi^0$ BR values on the $\alpha$ p.d.f., while the middle
plot combines the results from the analyses with the other
charmless hadronic $B$ decays. This combination gives
$\alpha = (92.2\pm 6.2)^{\circ}$.

Using the above inputs and our Bayesian framework, we perform
the global fit to extract the CKM matrix parameters $\rhob$
and $\etab$: we obtain $\rhob=0.137\pm0.022$ and
$\bar \eta=0.349\pm0.014$. The right plot in Figure~\ref{fig:sm} shows
the result of the SM fit on the $\bar\rho$-$\bar\eta$ plane.

With the default global fit, it is interesting to extract the
\utfit~predictions for SM observables. The BR$(B\to\tau\nu)$
is found to be $(0.81 \pm 0.07) \cdot 10^{-4}$ in agreement at the level
of $\sim 1.4 \sigma$ with the experimental measurement of
BR$(B\to\tau\nu) = (1.67 \pm 0.30) \cdot 10^{-4}$~\cite{hfag12}.
The BR$(B_s\to\mu\mu)$ is found to be $(3.88 \pm 0.15) \cdot 10^{-9}$,
while BR$(B^0\to\mu\mu)$ is $(1.13 \pm 0.07) \cdot 10^{-10}$
to be compared with the recent results by CMS and LHCb Collaborations
($\BR(B_s\to\mu\mu) = (2.8^{+0.7}_{-0.6}) \times 10^{-9}$ and
$\BR(B^0\to\mu\mu) = (3.9^{+1.6}_{-1.4}) \cdot 10^{-10}$~\cite{CMS:2014xfa}).

\section{Beyond the SM: Unitarity Triangle Analysis in presence of New Physics}
\label{sec:np}

We perform a full analysis of the UT reinterpreting the experimental
observables including possible model-independent NP contributions.
The possible NP effects considered in the analysis are those
entering neutral meson mixing ($\Delta F=2$ transitions) and they can
be parameterised in a model-independent way as:
\begin{eqnarray} 
  C_{B_q} \, e^{2 i \phi_{B_q}} & = & \frac{\langle
    B_q|H_\mathrm{eff}^\mathrm{full}|\bar{B}_q\rangle} {\langle
    B_q|H_\mathrm{eff}^\mathrm{SM}|\bar{B}_q\rangle}
  \;\; = \;\; \left(1+\frac{A_q^\mathrm{NP}}{A_q^\mathrm{SM}}
  e^{2 i (\phi_q^\mathrm{NP}-\phi_q^\mathrm{SM})}\right) \nonumber
\end{eqnarray}
where in the SM $C_{B_{d,s}}=1$ and $\phi_{B_{d,s}}=0$, or equivalently
$A_q^\mathrm{NP}=0$ and $\phi_q^\mathrm{NP}=0$. In addition,
$H_\mathrm{eff}^\mathrm{SM}$ is the SM $\Delta F=2$ effective Hamiltonian,
$H_\mathrm{eff}^\mathrm{full}$ is its extension in a general NP model, and
$q=d$ or $s$.

The following experimental inputs are added to the fit to extract information
on the $B_s$ system: the semileptonic asymmetry in $B_s$ decays,
the di-muon charge asymmetry, the $B_s$ lifetime from flavour-specific final
states, and CP-violating phase and the decay-width difference for $B_s$
mesons from the time-dependent angular analyses of $B_s\to J/\psi \phi$ decays.

\begin{figure}[!tb]
\centering
\vspace*{-1.6cm}
\hspace*{-0.6cm}
\includegraphics[width=0.36\linewidth]{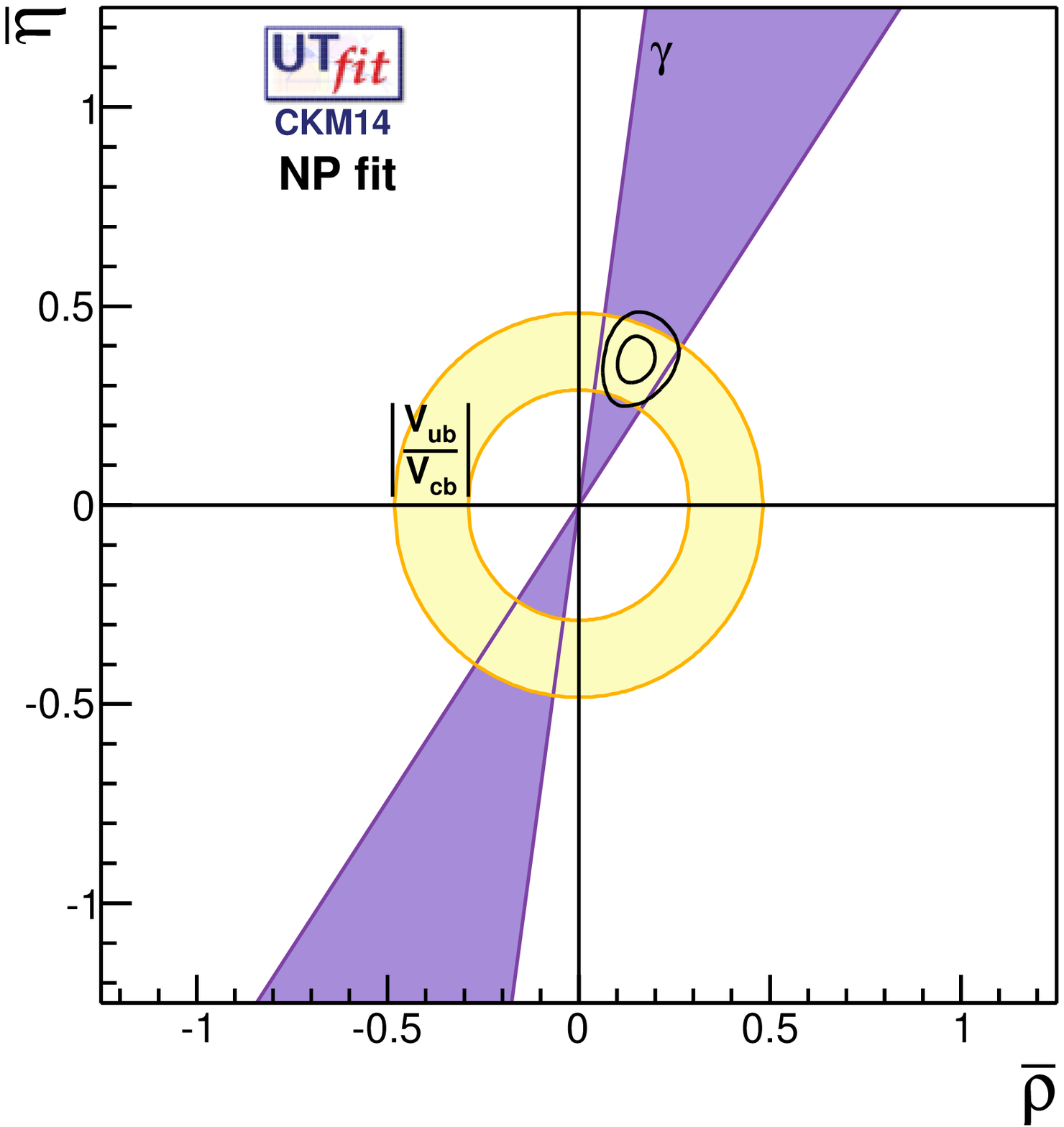}
\hspace*{-0.6cm}
\includegraphics[width=0.36\linewidth]{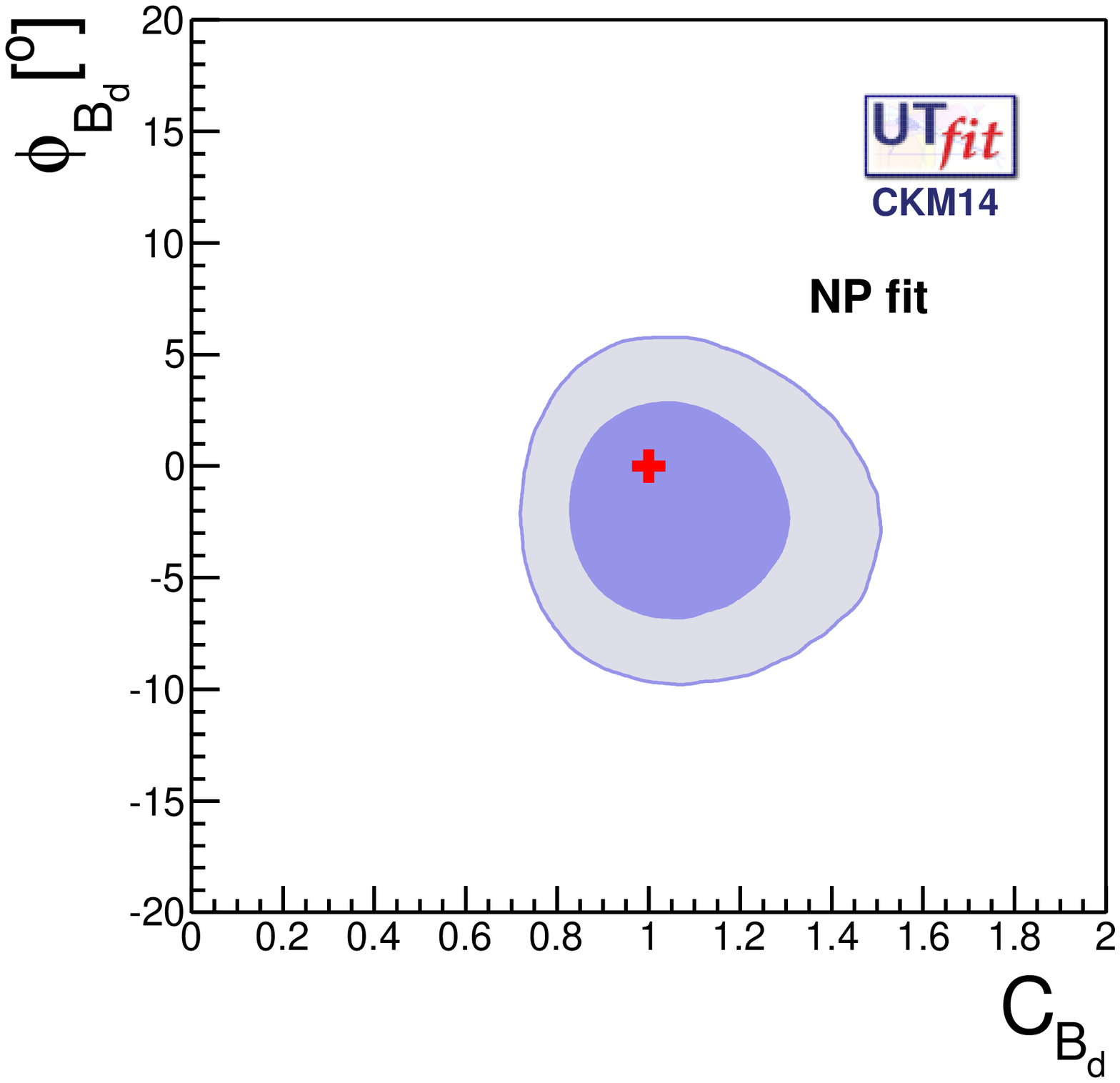}
\hspace*{-0.6cm}
\includegraphics[width=0.36\linewidth]{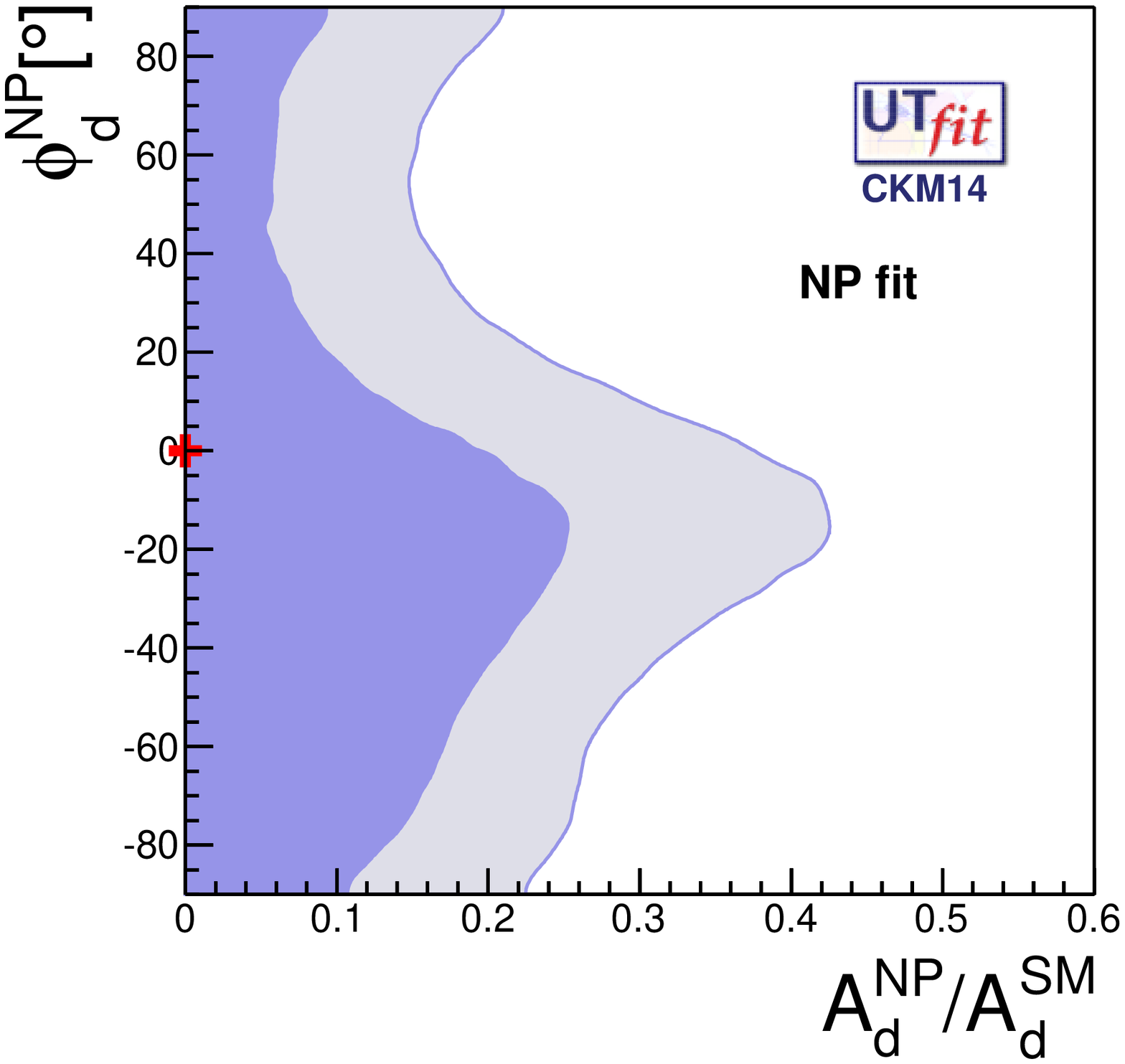}
\hspace*{-0.6cm}
\caption{{\it{Right}}: $\rhob-\etab$ plane showing the result
of the NP fit. The black contours display the 68\% and 95\% probability
regions selected by the given global fit. The 95\% probability
regions selected are also shown for those constraints
not affected by NP in $\Delta F=2$ transitions.
({\it{Middle-right}}) NP parameters in the $B_d$ system.
$68\%$ (dark) and $95\%$ (light) probability regions in the
$\phi_{B_d}$ -- $C_{B_d}$ ({\it{middle}}),
$\phi_{d}^\mathrm{NP}$ -- $\frac{A_d^{\mathrm{NP}}}{A_d^{\mathrm{SM}}}$ ({\it{right}})
planes in the NP fit. The red cross represents the SM expectation.}
\label{fig:np}
\end{figure}

From the full NP analysis, the global fit selects a region of the
$(\rhob, \etab)$ plane (left plot in Figure~\ref{fig:np},
with $\rhob=0.154\pm0.040$ and $\etab=0.367\pm0.048$)
which is consistent with the results of the SM analysis.
The NP parameters in the $B_d$ and $B_s$ systems are also extracted
from the fit and found in agreement with the SM expectations:
$C_{B_d}=0.81 \pm 0.12$, $\phi_{B_d}=(-3.4 \pm 3.6)^\circ$,
$C_{B_s}=0.87 \pm 0.09$ and $\phi_{B_s}=(-7 \pm 5)^\circ$.
The two right plots in Figure~\ref{fig:np} show the values still
available for the NP parameters in the $B_d$ system. Currently,
the ratio of NP/SM amplitudes needs to be less than $25\%$ at $68\%$
probability ($42\%$ at $95\%$ prob.) in $B_d$ mixing and less than $17\%$
at $68\%$ prob. ($25\%$ at $95\%$) in $B_s$ mixing.

We now consider the most general effective Hamiltonian for
$\Delta F=2$ processes ($\mathcal{H}_\mathrm{eff}^{\Delta F=2}$)
in order to translate the current constraints into allowed ranges
for the Wilson coefficients of $\mathcal{H}_\mathrm{eff}^{\Delta F=2}$.
The full procedure and analysis details are given in~\cite{Bona:2007vi}.
These coefficients have the general form
\begin{equation}
  \label{eq:cgenstruct}
  C_i (\Lambda) = \frac{F_i L_i}{\Lambda^2}\,
\end{equation}
where $F_i$ is a function of the (complex) NP flavour couplings, $L_i$
is a loop factor that is present in models with no tree-level Flavour
Changing Neutral Currents (FCNC), and $\Lambda$ is the scale of NP.
For a generic strongly-interacting theory with arbitrary flavour structure,
one expects $F_i \sim L_i \sim 1$ so that the allowed range for each
of the $C_i(\Lambda)$ can be immediately translated into a lower bound
on $\Lambda$.

\begin{figure}[!tb]
\centering
\vspace*{-1.6cm}
\hspace*{-0.2cm}
\includegraphics[width=0.36\linewidth]{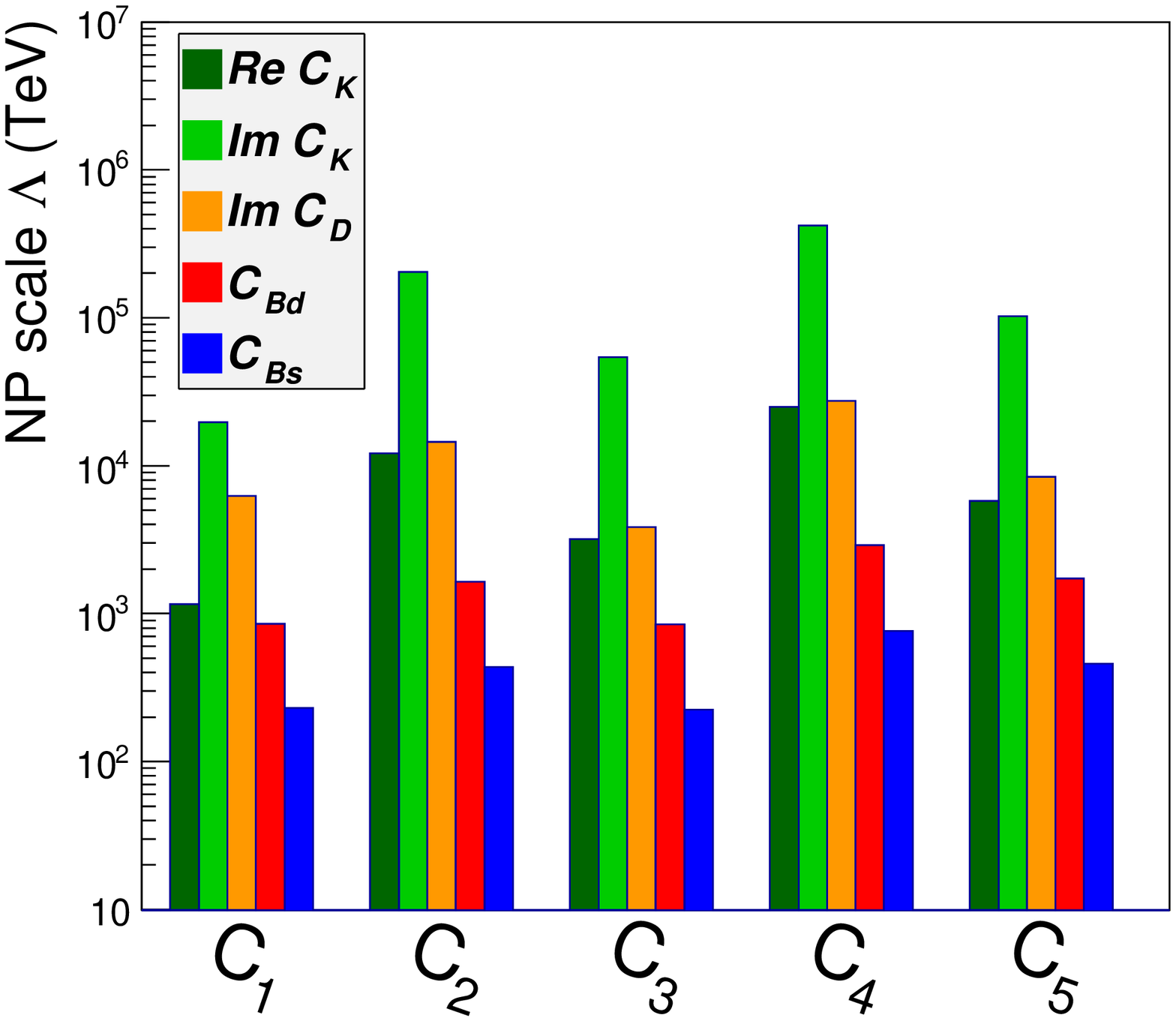}
\hspace*{-0.2cm}
\includegraphics[width=0.36\linewidth]{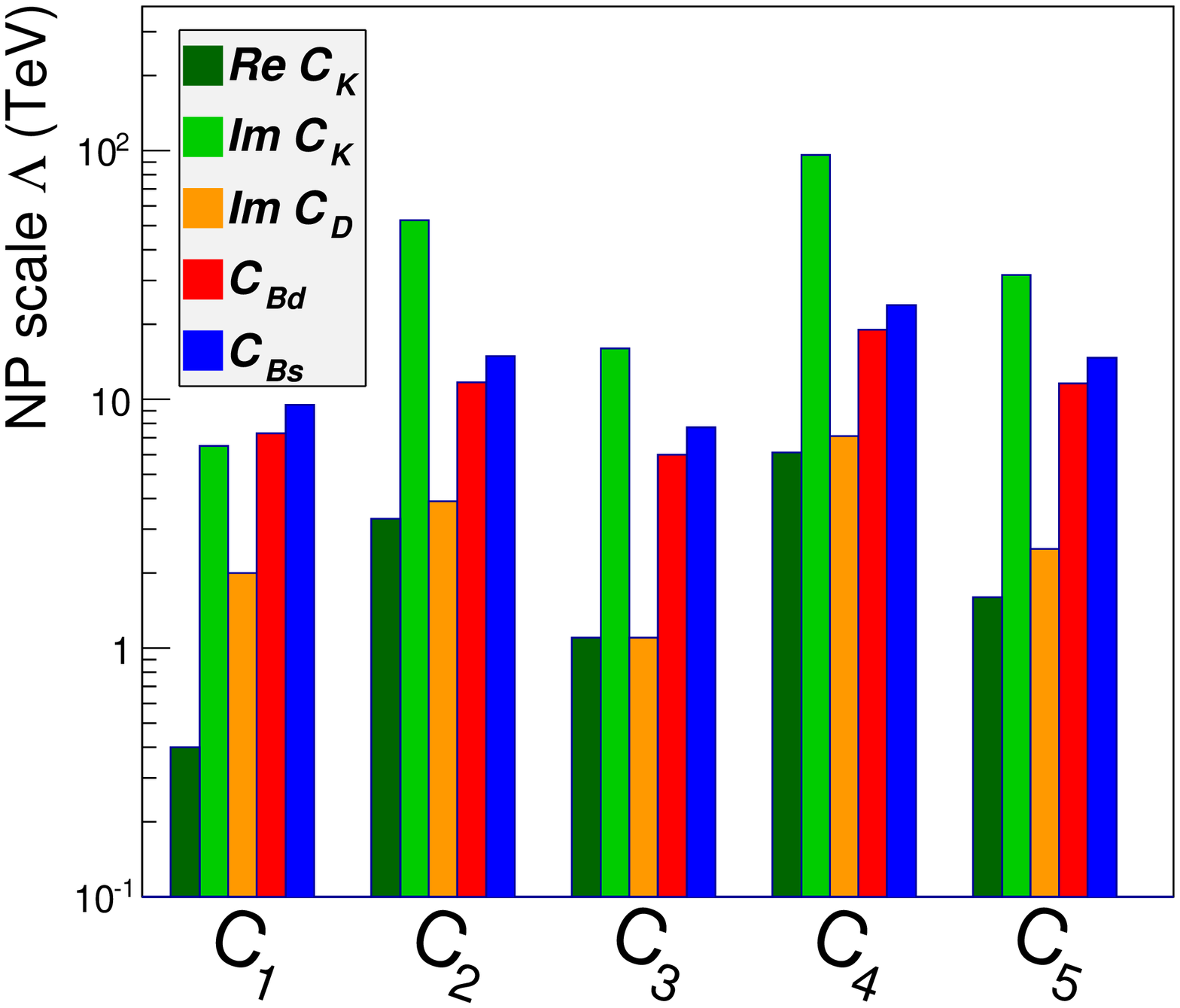}
\hspace*{-0.2cm}
\caption{Summary of the $95\%$ probability lower bound on the NP
scale $\Lambda$ for strongly-interacting NP in the general NP scenario
({\it{left}}) and in the NMFV scenario ({\it{right}}). Results from
all the neutral meson systems are shown.}
\label{fig:scales}
\end{figure}

The left plot in Figure~\ref{fig:scales} shows the results for the lower bounds
on $\Lambda$ coming from all the $C_i$'s for all the sectors in the case of
the general NP scenario, with arbitrary NP flavour structures
($\vert F_i \vert \sim 1$) with arbitrary phase and $L_i = 1$
corresponding to strongly-interacting and/or tree-level NP.
To obtain the lower bound on $\Lambda$ for loop-mediated
contributions, one simply multiplies the bounds we quote in the
following by $\alpha_s(\Lambda)\sim 0.1$ or by $\alpha_W \sim 0.03$.
The right plot in Figure~\ref{fig:scales} shows the lower bounds
on $\Lambda$ in a Next-to-Minimal-Flavour-Violation (NMFV) scenario
where the flavour structure is SM-like but with arbitrary phase relative
to the SM.

We conclude that any model with strongly interacting NP and/or
tree-level contributions is beyond the reach of direct searches at the
LHC, while in the case of weak couplings the lower bounds on the NP scale
are at the limit of the LHC reach.
The flavour sector provides the possibility of indirect searches
that remain a fundamental tool to constrain (or detect) NP at scales
higher that the LHC can provide.
\vspace*{0.4cm}

\bibliography{bona-utfit-ckm14}{}

\begin{thebibliography}{1}

\bibitem{Ciuchini:2000de}
M.~Ciuchini, G.~D'Agostini, E.~Franco, V.~Lubicz, G.~Martinelli, {\em et~al.},
  ``{\it{2000 CKM triangle analysis: A Critical review with updated
  experimental inputs and theoretical parameters}},'' {\em JHEP}, vol.~0107,
  p.~013, 2001.

\bibitem{Bona:2005vz}
M.~Bona {\em et~al.}, ``{\it{The 2004 UTfit collaboration report on the status
  of the unitarity triangle in the standard model}},'' {\em JHEP}, vol.~0507,
  p.~028, 2005.

\bibitem{hfag12}
Y.~Amhis {\em et~al.}, ``{Averages of B-Hadron, C-Hadron, and tau-lepton
  properties as of early 2012},'' 2012.

\bibitem{flag13}
S.~Aoki, Y.~Aoki, C.~Bernard, T.~Blum, G.~Colangelo, {\em et~al.}, ``{Review of
  lattice results concerning low-energy particle physics},'' {\em Eur.Phys.J.},
  vol.~C74, no.~9, p.~2890, 2014.

\bibitem{UTalpha}
M.~Bona {\em et~al.}, ``{Improved Determination of the CKM Angle alpha from B
  to pi pi decays},'' {\em Phys.Rev.}, vol.~D76, p.~014015, 2007.

\bibitem{bellepi0pi0}
\belle, ``{New Result on $B^0 \to \pi^0 \pi^0$ with Full data},'' 2014.
\newblock , presented at ICHEP 2014.

\bibitem{CMS:2014xfa}
V.~Khachatryan {\em et~al.}, ``{Observation of the rare $B^0_s\to\mu^+\mu^-$
  decay from the combined analysis of CMS and LHCb data},'' 2014.

\bibitem{Bona:2007vi}
M.~Bona {\em et~al.}, ``{\it{Model-independent constraints on $\Delta F=2$
  operators and the scale of new physics}},'' {\em JHEP}, vol.~0803, p.~049,
  2008.

\end{thebibliography}
\bibliographystyle{ieeetr}

\end{document}